\documentclass[reprint,superscriptaddress,amsmath,amssymb,aps,prl]{revtex4-1}
\pdfoutput=1

\usepackage{graphicx}
\usepackage{dcolumn}
\usepackage{bm}
\usepackage{hyperref}

\begin{document}

\title{Enhancing entropy and enthalpy fluctuations to drive crystallization in atomistic simulations}

\author{Pablo M. Piaggi}
\affiliation{Theory and Simulation of Materials (THEOS), {\'E}cole Polytechnique F{\'e}d{\'e}rale de Lausanne, c/o USI Campus, Via Giuseppe Buffi 13, CH-6900, Lugano, Switzerland}
\affiliation{Facolt{\`a} di Informatica, Instituto di Scienze Computazionali, and National Center for Computational Design and Discovery of Novel Materials MARVEL, Universit{\`a} della Svizzera italiana (USI), Via Giuseppe Buffi 13, CH-6900, Lugano, Switzerland}
\author{Omar Valsson}
\affiliation{Department of Chemistry and Applied Biosciences, ETH Zurich, c/o USI Campus, Via Giuseppe Buffi 13, CH-6900, Lugano, Switzerland}
\affiliation{Facolt{\`a} di Informatica, Instituto di Scienze Computazionali, and National Center for Computational Design and Discovery of Novel Materials MARVEL, Universit{\`a} della Svizzera italiana (USI), Via Giuseppe Buffi 13, CH-6900, Lugano, Switzerland}
\author{Michele Parrinello}%
 \email{parrinello@phys.chem.ethz.ch}
\affiliation{Department of Chemistry and Applied Biosciences, ETH Zurich, c/o USI Campus, Via Giuseppe Buffi 13, CH-6900, Lugano, Switzerland}
\affiliation{Facolt{\`a} di Informatica, Instituto di Scienze Computazionali, and National Center for Computational Design and Discovery of Novel Materials MARVEL, Universit{\`a} della Svizzera italiana (USI), Via Giuseppe Buffi 13, CH-6900, Lugano, Switzerland}
\date{\today}

\begin{abstract}
Crystallization is a process of great practical relevance in which rare but crucial fluctuations lead to the formation of a solid phase starting from the liquid.
Like in all first order first transitions there is an interplay between enthalpy and entropy.
Based on this idea, in order to drive crystallization in molecular simulations, we introduce two collective variables, one enthalpic and the other entropic.
Defined in this way, these collective variables do not prejudge the structure the system is going to crystallize into.
We show the usefulness of this approach by studying the case of sodium and aluminum that crystallize in the bcc and fcc crystalline structure, respectively.
Using these two generic collective variables, we perform variationally enhanced sampling and well tempered metadynamics simulations, and find that the systems transform spontaneously and reversibly between the liquid and the solid phases.
\begin{description}
\item[PACS numbers]
05.10.-a, 02.70.Ns, 64.70.D-
\end{description}
\end{abstract}

\maketitle

Crystallization is a remarkable physical process in which the disordered atoms of a liquid spontaneously form beautifully ordered periodic patterns.
It is also a phenomenon of great importance in many areas of application, from metallurgy to material science, pharmacology and even biology.
Not surprisingly this problem has received considerable attention.
Understanding the way in which crystallization proceeds holds also the key to improve many scientific and technological processes.
It suffices to recall here the difficult art of protein crystallization.
Since experiments can only provide a limited insight into this phenomenon, already in the very early days of computer simulation much effort has been devoted to the study of crystal nucleation \cite{Mandell76,Stillinger78,Hsu79}.

Unfortunately in most cases the time scale of nucleation is much longer than what can be reached in an atomistic simulation.
Early on this problem was tackled by forcing nucleation using unphysically deep temperature quenches so as to bring the nucleation time scale within the reach of simulation \cite{Hsu79}.
This procedure is not without problems since it can even change the nature of the nucleation process \cite{Trudu06}.
For this reason enhanced sampling methods have been extensively used \cite{vanDuijneveldt92,Moroni05,Salvalaglio14}.
Most of them are based on the definition of appropriate collective variables (CVs) able to distinguish one type of local order from another.
Typical examples of CVs used in this context are the Steinhardt order parameters \cite{Steinhardt83,vanDuijneveldt92} or the ones introduced by Santiso and Trout \cite{Santiso11,Salvalaglio14, Giberti15}.

However, the use of these kind of CVs can prejudge the structure the system is going to crystallize into.
Thus, it would  be extremely useful to have an enhanced sampling method that does not assume from the start the final structure.
Such a method could illuminate important details of nucleation and complement structure prediction methods that are based on finding the crystal structure of lowest energy \cite{Goedecker04,Oganov06}.
This would be extremely valuable in all those cases in which the crystal structure is stabilized by strong entropic effects as in superionic or plastic crystals.

\begin{figure*}
	\begin{center}
		\includegraphics[width=0.99\textwidth]{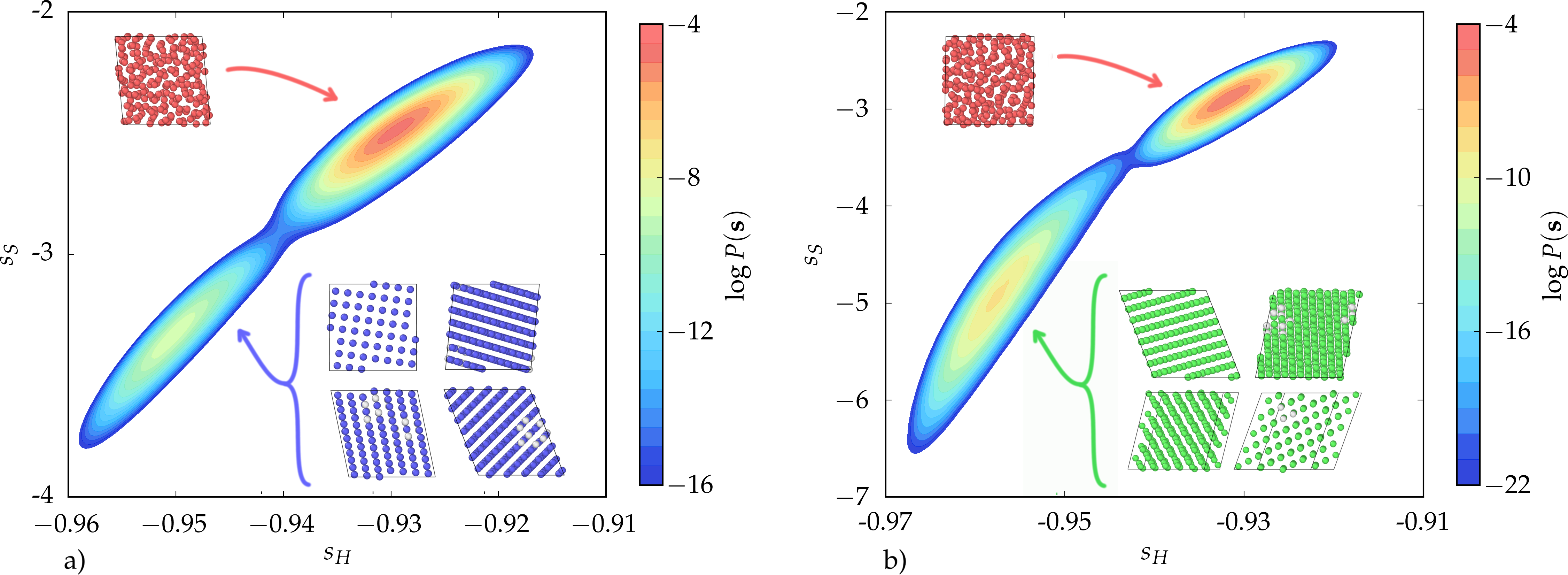}
		\caption{\label{fig:Figure1} Marginal probability distribution with respect to $s_H$ and $s_S$.
    a) Na at 350 K.
    b) Al at 800 K.
    A liquid and a solid basin are observed in the plots and some characteristic configurations in each basin are depicted.
    The solid structures were identified using CNA.
    Bcc-like atoms are colored in blue and fcc-like atoms are colored in green.
    As expected, the solid configurations in Na and Al, have a bcc and fcc crystalline structure, repectively.
    A liquid configuration is also shown with liquid-like atoms colored in red.
    $s_H$ is expressed in units of the cohesive energy and $s_S$ is in $k_B$.
    }
	\end{center}
\end{figure*}

In order to achieve this result the CV should not be related to this or that feature of the geometry of the crystal.
It thus comes natural to recall that in crystallization, like in all first order transformations, there is a trade off between enthalpy and entropy, and introduce two CVs able to describe these changes.
Of course enthalpy $H=E+PV$ where $E$ is the total energy, $P$ the pressure and $V$ the volume, is easy to estimate while for the entropy there is no exact expression.
However in order to bias the system we do not need to compute the exact entropy and even an approximate expression will do.
The liquid state theory provides an expression in which the excess entropy per atom is expanded in an infinite series of terms involving multiparticle correlation functions \cite{Nettleton58}.
The first such term $S_2$, used in ref.\ \citenum{Giuffre10, Truskett00, Zhang11} to perform insightful analysis, includes only two body correlations and is given by:
\begin{equation}
	S_2=-2\pi\rho k_B \int\limits_0^{\infty} \left [ g(r) \ln g(r) - g(r) + 1 \right ] r^2 dr .
	\label{eq:pair_entropy}
\end{equation}
In this equation $g(r)$ is the radial distribution function and $\rho$ is the density of the system.
We shall use $S_2$ as CV.
In a different context the use of $S_2$ has already been suggested to address rare events problems in which the entropy plays a dominant role \cite{Zhou06}.
Both $H$ and $S_2$ have a proper thermodynamic meaning only when averaged.
However, for the purpose of this paper we use their instantaneous values to define useful CVs.

In this spirit we introduce two CVs, one enthalpic $s_H$ and the other entropic $s_S$.
The former is defined as:
\begin{equation}
  s_H = \frac{U(\mathbf{R}) + PV}{N},
\end{equation}
where $U(\mathbf{R})$ is the potential energy and $N$ the number of atoms in the system \cite{Bonomi10}.
We thus do not include the kinetic energy contribution to the total energy in the definition of $s_H$.
The latter is slightly more complex.
We first define a mollified version of the radial distribution function:
\begin{equation}
	g_m(r) = \frac{1}{4 \pi N \rho r^2} \sum\limits_{i \neq j} \frac{1}{\sqrt{2 \pi \sigma^2}} e^{-(r-r_{ij})^2/(2\sigma^2)} ,
  \label{eq:mollified_rdf}
\end{equation}
where $r_{ij}$ is the distance between particles $i$ and $j$, and $\sigma$ is a broadening parameter.
The mollification is necessary to ensure that the derivatives of $g_m(r)$ relative to the atomic positions are continuous.
The resulting $g_m(r)$ is inserted into equation \eqref{eq:pair_entropy} and the integral calculated numerically using the trapezoid rule up to a cut off distance $r_{\mathrm{max}}$.
In the spirit of the work of Tiwary and Berne \cite{Tiwary16} $r_{\mathrm{max}}$ is chosen so as to optimize the frequency of transitions between solid and liquid.

Before describing the details of our simulation we would like to pause a little to discuss a small but crucial technical issue.
It has long been recognized that a given crystalline order can conflict with the periodic boundary conditions of molecular dynamics (MD) simulations and that allowing the MD cell shape to vary is essential to study processes that involve crystals.
This is at the heart of the Parrinello-Rahman method that allows the MD cell to change under conditions of constant stress \cite{Parrinello81}.
The MD cell shape is expressed by an upper diagonal matrix $ \mathbf{h} = [h_{11}, h_{22}, h_{33}, h_{23} , h_{13}, h_{12}] $.
Changes in $\mathbf{h}$ are driven by the unbalance between the internal and the external stress.
A straightforward application of the Parrinello-Rahman method to the liquid state is however problematic since a liquid offers no resistance to shear and left to its own devices the MD box shape would fluctuate randomly assuming even inconvenient cigar like shapes.
In order to remedy this we take $h_{11}=h_{22}=h_{33}$ and this forces the volume to be always close to a cube but still allows the off diagonal elements to vary, and thus accomodate different structures.

We now exemplify the usefulness of this approach in the cases of sodium and aluminum that crystallize in the bcc and fcc structures, respectively.
We simulated Na and Al using embedded atom models \cite{Wilson15, Mendelev08} whose melting temperatures have been determined.
Biased MD simulations were performed using LAMMPS \cite{Plimpton95} patched with PLUMED 2 \cite{Tribello14} and the VES code \cite{vescode}.
The integration of the equations of motion was carried out with a timestep of 2 fs.
We employed the stochastic velocity rescaling thermostat \cite{Bussi07} with a relaxation time of 0.1 ps.
The target pressure of the barostat was set to its standard atmospheric value and a relaxation time of 10 ps was used.
Systems composed of 250 and 256 atoms were employed for Na and Al, respectively.
In order to determine the Gibbs free energy surface $G(\mathbf{s})$ as a function of $s_H$ and $s_S$ we used well tempered metadynamics \cite{Barducci08} and the variationally enhanced sampling (VES) \cite{Valsson14} method in its well tempered variant \cite{Valsson15}.
Details of the calculations can be found in the Supplemental Material.
For the calculation of $s_S$ we used a cutoff of $r_{\mathrm{max}}=0.65$ nm for Na and $r_{\mathrm{max}}=0.70$ for Al.
The broadening parameter $\sigma$ in equation \eqref{eq:mollified_rdf} and the step size in the numerical integration of equation \eqref{eq:pair_entropy} was $0.0125$ nm both for Na and Al.

As shown in Fig.\ \ref{fig:Figure1}, employing these CVs we were able to describe accurately the phase transition between the liquid and crystalline phases in Na and Al without having had to feed any prior information on the systems.
In both cases two basins were observed, one of high enthalpy and high entropy, and another of low enthalpy and low entropy, corresponding to the liquid and solid basins, respectively.
This reflects the trade off discussed in the introduction between enthalpy and entropy in first order phase transitions.
As hoped for, Na crystallized into the bcc structure whereas Al crystallized in the fcc structure.
In all cases the transition is reversible (see Supplemental Material) allowing the free energy surface to be estimated.

We performed a detailed analysis of the crystalline structures in the trajectories using common neighbor analysis (CNA) \cite{Stukowski09,Honeycutt87}.
For Na the only solid phase that can be identified in the simulation is bcc.
In the case of Na either $s_S$ or $s_H$ can act as order parameters, since any of them by itself could be able to distinguish between the liquid and bcc phases.
However they alone do not fully capture the nature of the transition.
We have performed simulations biasing only $s_S$ or $s_H$ and we found that their efficiency in metadynamics is very low.
Instead if they are biased together the sampling efficiency is greatly improved.
Furthermore, the FES is essentially one dimensional thus no extra computational burden is posed by the use of two CVs rather than one.
For Al the situation is slightly more complex since there is a bcc phase at high free energies.
In this case the use of both $s_S$ and $s_H$ is necessary to distinguish between the liquid, fcc, and bcc phases.
For compact structures such as fcc, the formation of stacking faults must be considered.
In our simulations these defects rarely form since the stacking fault energy of the Al potential \cite{Mendelev08} reproduces well the high experimental value.
These results are discussed to a greater detail in the Supplemental Material.

\begin{figure}
        \begin{center}
                \includegraphics[width=0.5\textwidth]{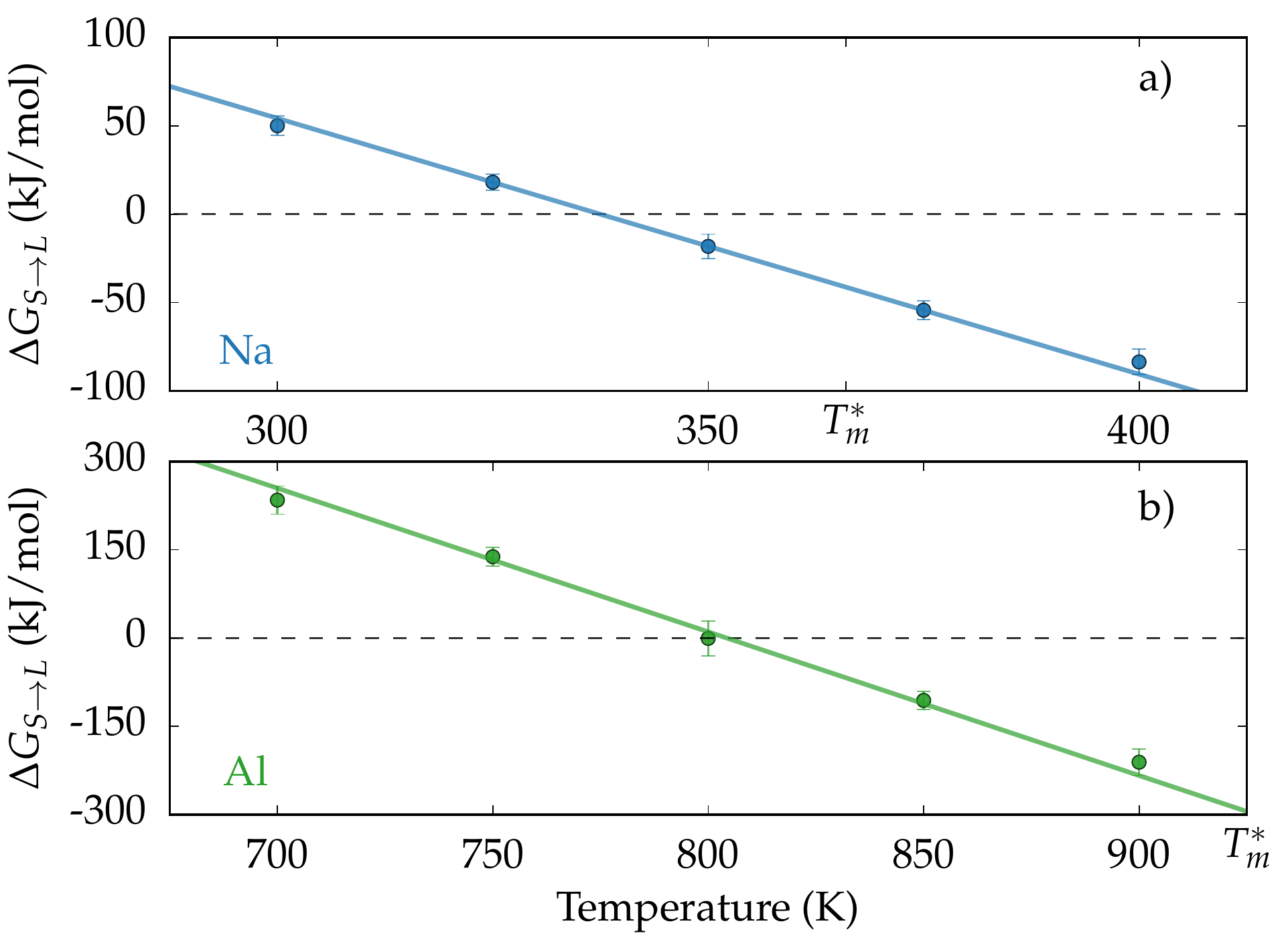}
                \caption{\label{fig:Figure2} Difference in free energy between the liquid and solid phases as a function of temperature as calculated from eq.\ \eqref{eq:dif_free_energy}.
    Subplots a) and b) correspond to Na and Al, respectively.
    The melting temperatures reported in ref.\ \citenum{Wilson15} and \citenum{Mendelev08} are marked in the abscissa axis with $T_m^*$.
    The straight lines have only been fitted to the three middle points.
    }
        \end{center}
\end{figure}

From our calculation we can also get the difference in free energy between the two phases as:
\begin{equation}
	\Delta G_{S \rightarrow L} = -\frac{1}{\beta} \log \left ( \frac{\int_{\mathrm{L}} d\mathbf{s} e^{-\beta G(\mathbf{s})}}{ \int_{\mathrm{S}} d\mathbf{s} e^{-\beta G(\mathbf{s})} } \right)
  \label{eq:dif_free_energy}
\end{equation}
where $\mathbf{s}$ is the set of CVs $s_H$ and $s_S$ and the integrals are restricted to the liquid (L) and solid (S) basins, respectively.
In Fig.\ \ref{fig:Figure2}a we plot $\Delta G_{S \rightarrow L}$ for Na in the temperature range 300-400 K.
From this calculations, we estimate the melting temperature to be around 340 K in agreement with results obtained using a different technique (366 K) \cite{Wilson15}, if one takes into account that the melting temperature is a strong function of system size \cite{Sturgeon00}.
For details of these calculations we refer the reader to the Supplemental Material.
In Fig.\ \ref{fig:Figure2}b we plot $\Delta G_{S \rightarrow L}$ for Al in the temperature range 700-900 K.
The melting temperature is around 800 K, somewhat below the melting temperature calculated from coexistence simulations (926 K).
Again this is due to strong finite size effects.

Once that $\Delta G_{S \rightarrow L}$ is determined as a function of temperature, the difference in entropy between the liquid, $\Delta S_{S \rightarrow L}$, can be calculated from the thermodynamic identity $\Delta S_{S \rightarrow L}=-\frac{\partial \Delta G_{S \rightarrow L}}{\partial T}\Bigr|_{N,P}$.
Since also $\Delta H_{S \rightarrow L}$ can be calculated, the definition of Gibbs free energy $\Delta G_{S \rightarrow L} = \Delta H_{S \rightarrow L} - T \Delta S_{S \rightarrow L}$ provides an independent estimate of $\Delta S_{S \rightarrow L}$.
In Table \ref{tab:table1} we compare these two estimates of the entropy and calculations from experiments, both for Na and Al.
\begin{table}[b]
\caption{\label{tab:table1}%
Difference in entropy between the solid and liquid phases $\Delta S_{S \rightarrow L}$ at the melting temperature.
$\Delta S_{S \rightarrow L}$ was calculated using two different approaches and the results are compared with the experimental value.
}
\begin{ruledtabular}
\begin{tabular}{cccc}
 &
$-\frac{\partial \Delta G_{S \rightarrow L}}{\partial T}\Bigr|_{N,P}$ \footnote{The unit is J  K$^{-1}$ mol$^{-1}$} &
$ \frac{\Delta H_{S \rightarrow L} - \Delta G_{S \rightarrow L} }{T}$ &
\textrm{Experimental \footnote{Ref.\ \cite{Chase86} } } \\
\colrule
\textrm{Na} & 5.8 & 6.6 & 7.017\\
\textrm{Al} & 9.5 & 10.7 & 11.475 \\
\end{tabular}
\end{ruledtabular}
\end{table}
The estimates are comfortably similar and in line with the experimental values \cite{Chase86}.

In conclusion, the use of a collective variable that couples directly to entropy has proven to be very promising.
We have illustrated the power of the approach by crystallizing Na and Al in their minimum free energy structures.
The success of our calculation suggests a general strategy for tackling those problems in which entropy alone or in combination with enthalpy plays a role.
This in the practice means that one only needs to find approximate ways of expressing the entropy and the enthalpy.
Once that the variables have been chosen, the use of metadynamics \cite{Laio02} or VES can amplify the CV fluctuations and accelerate the observation of the desired transition.
We stress that a rigorous definition of entropy is not necessary and metadynamics or VES are very accommodating in this respect.
We can anticipate here that the strategy of biasing an entropic CV is being adapted to the folding of small proteins.

\section*{Acknowledgements}

This research was supported by the NCCR MARVEL funded by the Swiss National Science Foundation.
The authors also acknowledge funding from the European Union Grant No. ERC-2014-AdG-670227 / VARMET.
The computational time for this work was provided by the Swiss National Supercomputing Center (CSCS).
Calculations were performed in CSCS clusters Piz Daint and M{\"o}nch.
We are grateful to Paolo V. Giaquinta and David Chandler for useful discussions.

\end{document}